\documentclass[a4paper,UKenglish,cleveref, autoref, thm-restate]{oasics-v2021}
\pdfoutput=1

\usepackage{listings,xfp}
\usepackage{anyfontsize}
\usepackage{etex} %avoid error messages from tikz due to bug
\usepackage{tikz}

\usepackage{mdwlist} % compact list itemize* enumerate*
\newcommand\deffont[1]{\textbf{#1}}
\newcommand\entrypoint[1]{\lstinline{\%#1}}

\usepackage{microtype}
\usepackage[T1]{fontenc}
\usepackage[scaled]{beramono}
\newcommand\Small{\fontsize{9}{9.2}\selectfont}
\newcommand\Tiny{\fontsize{7}{7.2}\selectfont}
\newcommand*\LSTfont{\Small\ttfamily\SetTracking{encoding=*}{-60}\lsstyle}
\newcommand*\LSTfonttiny{\Tiny\ttfamily\SetTracking{encoding=*}{-60}\lsstyle}

\hypersetup{colorlinks = true}
\title{Money grows on (proof-)trees: the formal FA1.2 ledger standard} 

\titlerunning{The formal FA1.2 ledger standard} 

\author{Murdoch J. Gabbay}{Heriot-Watt University, Edinburgh, UK and \\ Nomadic Labs, Paris, France \and \url{http://www.gabbay.org.uk} }{}{https://orcid.org/0000-0001-5796-3455}{}

\author{Arvid Jakobsson}{Nomadic Labs, Paris, France \and \url{https://www.arvidj.eu}}{}{https://orcid.org/0000-0001-6028-8972}{}

\author{Kristina Sojakova\footnote{Sojakova wrote the Coq code described in this paper}}{INRIA, Paris, France \and \url{https://who.rocq.inria.fr/Kristina.Sojakova/}}{}{}{}

\authorrunning{M. J. Gabbay, A. Jakobsson, and K. Sojakova} %TODO mandatory. First: Use abbreviated first/middle names. Second (only in severe cases): Use first author plus 'et al.'

\Copyright{CC-BY \url{http://creativecommons.org/licenses/by/3.0/}} %TODO mandatory, please use full first names. OASIcs license is "CC-BY";  http://creativecommons.org/licenses/by/3.0/

\ccsdesc[500]{Software and its engineering~Formal software verification}
\ccsdesc[300]{Theory of computation~Program verification}
\ccsdesc[300]{Social and professional topics~Quality assurance}

\keywords{Distributed ledger, smart contracts, Coq, formal verification, blockchain} %TODO mandatory; please add comma-separated list of keywords

\acknowledgements{The authors acknowledge the support of Nomadic Labs and the detailed and very helpful comments of three anonymous referees.  We also thank Benoît Rognier, the Edukera team, and Tom Jack for their feedback and support.}

\nolinenumbers %uncomment to disable line numbering

\EventShortTitle{FMBC 2021}
\EventAcronym{FMBC}
\EventYear{2021}
\begin{document}

\maketitle

\begin{abstract}
Once you have invented digital money, you may need a ledger to track who owns what --- along with an interface to that ledger so that users of your money can transact.
On the Tezos blockchain this implies: a smart contract (distributed program), storing in its state a ledger to map owner addresses
to token quantities; along with standardised entrypoints to query and transact on accounts.

A bank does a similar job --- it maps account numbers to account quantities and permits users to transact --- but in return the bank demands trust, it incurs expense to maintain a centralised server and staff, it uses a proprietary interface \dots and it may speculate using your money and/or display rent-seeking behaviour.
A blockchain ledger is by design decentralised, inexpensive, open, and it won't just decide to bet your tokens on risky derivatives (unless you want it to). 

The FA1.2 standard is an open standard for ledger-keeping smart contracts on the Tezos blockchain.
Several FA1.2 implementations already exist.

Or do they?
Is the standard sensible and complete?
Are the implementations correct?
And what are they implementations \emph{of}?
The FA1.2 standard is written in English, a specification language favoured by wet human brains but notorious for its incompleteness and ambiguity when rendered into dry and unforgiving code.

In this paper we report on a formalisation of the FA1.2 standard as a Coq specification, and on a formal verification of three FA1.2-compliant smart contracts with respect to that specification.
Errors were found and ambiguities were resolved; but also, there now exists a \emph{mathematically precise} and battle-tested specification of the FA1.2 ledger standard.

We will describe FA1.2 itself, outline the structure of the Coq theories --- which in itself captures some non-trivial and novel design decisions of the development --- and review the detailed verification of the implementations.  
\end{abstract}

\section{Introduction}
\label{sect.intro}

\subsection{Tezos: a universal, modular blockchain}

The \href{https://en.wikipedia.org/wiki/Tezos}{Tezos blockchain} was outlined in a 2015 whitepaper~\cite{tezos-white-paper} and went live in September 2018.
It is an accounts-based proof-of-stake blockchain system with the unique feature that it is a \emph{universal blockchain} in the sense that the protocol for running Tezos is itself data on the Tezos blockchain,
and this data is subject to regular upgrade by stake-weighted community vote.\footnote{
As the programs of the `universal' Turing machine are themselves data on its memory.
The `regular upgrade' property is called a \emph{self-amendment} in the Tezos literature.

To be precise: for space-efficiency, the Tezos blockchain holds not the protocol code but its hash --- it is a standard trick to store large datastructures off-chain and retain an on-chain hash. 
When the protocol self-amends the hash gets updated and code matching that hash --- which must be the protocol code (where `must' = `our hash function is computationally infeasible to break') --- propagates across the network for nodes to load and run.  This low-level functionality is handled by a `shell' (think: BIOS).
}
Universality favours a healthy modularity at every level of the system's design, since almost anything in the running system can be and is subject to update.

Tezos has \emph{just one} native token: the \emph{tez}.
Further tokens can be created in a modular fashion, using smart contracts.

Thus we can represent Ethereum and Bitcoin on Tezos (using so-called \emph{wrapped tokens});\footnote{We mention a few wrapped tokens at the start of Section~\ref{sect.refining}.}
 we can represent NFTs (non-fungible tokens representing unique assets); likewise for stablecoins and so forth.
All these things can be and have been represented 
as Tezos smart contracts. 
Given this freedom, we need \emph{interoperability standards} for our tokens to adhere to.
After all, a token on its own is useless; its value comes from how we might \emph{transact} with it.\footnote{Like money in the bank is only useful because you could use it to perform transactions.  You don't \emph{have} to --- at least not all at once --- but that's not the point: what matters is that you \emph{could}.}

%%%%%%%%%%%%%%%%%%%%%%%%%%%%%%%%
\subsection{The FA1.2 standard: five entrypoints, in English}
\label{subsect.five.entrypoints}

The \deffont{\href{https://tzip.tezosagora.org/proposal/tzip-7/}{FA1.2 standard}} is an English document specifying a minimal API for a ledger-like smart contract.
Compliance with FA1.2 ensures some degree of interoperability across multiple smart contracts and tools on the Tezos blockchain.

The FA1.2 standard asserts that a given smart contract should provide the following five entrypoints and behaviours:
\begin{enumerate}
\item
\entrypoint{transfer} expects a \deffont{from} account, a \deffont{to} account, and an \deffont{amount} of tokens to be transferred, and updates the ledger accordingly.
\item
\entrypoint{approve} expects an \deffont{owner}, a \deffont{spender}, and a \deffont{new allowance} for the spender, and updates the transfer approvals accordingly.\footnote{When you use a debit card you \emph{authorise} a debit.  The merchant could in principle not do this; thus the authorisation is granted but the withdrawal does not take place.
Likewise a direct debit is an approval for a withdrawal.
Similarly \entrypoint{approve} does not send tokens; it approves another smart contract to make a token withdrawal, up to some limit.  E.g. when you sell tokens for tez in Dexter, you give permission using \entrypoint{approve} for Dexter to transfer tokens from your account.}
\item
\entrypoint{getAllowance} expects an \deffont{owner}, a \deffont{spender}, and returns the approved transfer allowance for the spender, via a callback (see Remark~\ref{rmrk.pass.tokens} below).
\item
\entrypoint{getBalance} expects an \deffont{owner} and returns the owner's balance via a callback.
\item\label{item.getTotalSupply}
\entrypoint{getTotalSupply} returns the total sum of all balances in the ledger, via a callback.
\end{enumerate}

\begin{remark}
\label{rmrk.left.out}
The list above is nearly a complete summary of the FA1.2 standard, which is just a couple of pages long and clearly intended to be as straightforward as possible (which is a good thing).
A few words may be helpful on what this standard \emph{leaves out}:
\begin{enumerate}
\item
The FA1.2 standard \emph{does not} exclude entrypoints that are not mentioned in the list above: an implementation may offer additional entrypoints. 
\item
The FA1.2 standard \emph{does not} constrain the behaviour of additional entrypoints, if present in an implementation.
These entrypoints could change balances or allowances (such as some kind of \emph{admin} entrypoint) or the total number of tokens in circulation (often called \emph{mint} or \emph{burn} entrypoints).
\item 
The FA1.2 standard \emph{does not} exclude additional preconditions on the entrypoints that it mentions:
an implementation may impose additional preconditions. 
\end{enumerate}
For example, consider a simple standard for doors that insists on just one entrypoint, \entrypoint{openDoor}, with two conditions: if the call to \entrypoint{openDoor} succeeds then the door will be open afterwards; and the call fails with error `alreadyOpen' if the door is already open. 
Hopefully your home's front door complies with this standard, which is why it is called \emph{a door}, but also: it has an additional implementation-specific entrypoint \entrypoint{lockDoor}; and an additional implementation-specific precondition on its \entrypoint{openDoor} entrypoint that it will fail --- even if the door is not open! --- if the door is locked; and finally the door has an administrative override entrypoint \entrypoint{fireDepartment}, to be invoked only by people with authorisation and in special circumstances.\footnote{True story.  The first author has seen this entrypoint invoked.}
Yet none of this implementation-specific structure makes your front door any less of \emph{a door}. 
More on this is in Section~\ref{sect.refining} and Remark~\ref{remark.not.in}.
\end{remark}

\begin{remark}[Callbacks]
\label{rmrk.pass.tokens} 
A call to any entrypoint of a smart contract in Tezos takes some parameters, some (possibly zero) quantity of tez, and a continuation address of another entrypoint, called a \emph{callback}, to which flow of control will continue. 
Thus ``returns X via a callback'' above means X is passed as a parameter to the nominated callback entrypoint.
\end{remark}

\subsection{This is not enough}
\label{subsect.not.enough}

The English FA1.2 standard is reasonable \emph{per se}, but it is not enough:
\begin{enumerate}
\item\label{bloody-minded}
The FA1.2 standard is written in English.
This means it \emph{might} be incomplete or incoherent,\footnote{In fact there's no `might' about it: a quick scan of the standard reveals points which a suitably na\"ive, bloody-minded, or hostile reader could interpret in spectacularly different ways, in spite of the authors' efforts to be precise.  Thus multiple implementations could exist, doing radically different things and all claiming plausibly to be `true' to `the' FA1.2 standard.  This is not a criticism of FA1.2 or its authors: it is in the nature of the English language itself.}
 and it \emph{can't} be directly manipulated using verification tools.
\item
Just because a smart contract claims to be FA1.2-compliant does not mean that it is: perhaps it is buggy; perhaps it is hostile; perhaps the implementors just interpreted the English specification differently than the standard's authors intended. 
\item\label{standard.standard}
The FA1.2 standard is not itself a standard for verifying compatibility with the FA1.2 standard!
That is: given two verifications of two implementations (or even of the same implementation), it is not \emph{a priori} guaranteed that they are verifying \emph{the same properties} --- and the FA1.2 standard, which is written in English, cannot help resolve this.
\end{enumerate}
To state the obvious: ledgers are safety-critical.
This is real money --- for a certain 21st century definition of `real' --- that our smart contracts could be manipulating~\cite{nikolic:fingps,atzei:suraes}.

Saying `trust us, we're experts' is problematic not just because we might be wrong, but also because an open permissionless blockchain should not demand such trust: users should be able to check correctness, or trust that somebody independent of a central `expert' authority has checked or could check this, and (since this is an open system) they should best also trust that whatever `correctness' means, it means nearly, and preferably precisely, the same to them as to the other users with whom they might transact.

\subsection{Our work in a nutshell}

This paper reports on a verification effort undertaken at Nomadic Labs that we argue addresses the points above.
That is: 
\begin{itemize*}
\item
we place the FA1.2 standard on a precise mathematical footing that can be both trusted and verified, and
\item
we check correctness of three smart contracts which claim to be FA1.2-compliant.
\end{itemize*}
The reader should not expect novel maths in this work --- indeed, in this context `novel' = `untested' and may best be avoided where possible.
However, there are other types of innovation to this work: 
\begin{enumerate*}
\item
To our knowledge, this is the only full formalisation of a blockchain ledger standard and of multiple implementations against it, in a theorem-prover.

This addresses the three points above, by providing: a formal specification of the standard, formal representations within the theorem-prover of the programs themselves, proofs of compliance for the latter with respect to the former --- and also a gold standard for comparing and operating on all of these proofs, since they are all proof-objects within the theorem-prover itself.
\item 
Also relevant is the theory files' structure, which is new as we discuss below.
\end{enumerate*}
Having secure and reliable ledgers on Tezos is an existential issue for the blockchain ecosystem, so the fact that this could be nailed down, as we have done, has both practical and theoretical importance.
Thus, this work exemplifies the application to a tangible industrial problem of a particular (Coq-based) theorem-prover technology ecosystem.

\begin{remark}
We may write \emph{smart contract} and \emph{implementation} synonymously. 
Also, smart contracts may be written in high-level languages, but to run on Tezos they must get compiled to a lower-level stack-based language called Michelson.\footnote{Think: the Tezos equivalent of bytecode or machine code, though Michelson is still quite high level.}
We may not always distinguish between the original program and its compiled Michelson executable, but we will when this difference matters and it will always be clear what is meant.
\end{remark}

\begin{remark}
The formal FA1.2 standard does not replace the English FA1.2 standard: to be fully proficient a reader would have to know Coq \emph{and also} understand something about how Tezos contracts are embedded into it.
However, the formal FA1.2 standard (or an evolution of it) \emph{could} serve as a standard reference within certain expert communities, and even outside such communities a reader with good reason to try could parse the Coq code, if they have some knowledge of dependent types and perhaps have read the English standard and looked at this paper.
Thus, the formal and the English FA1.2 standards are embedded in what one might call a larger `space of understanding', within which they complement and enrich one another such that each is made stronger and more rigorous by the existence of the other.
\end{remark}

%%%%%%%%%%%%%%%%%%%
\section{Introducing: the formal FA1.2 standard}
\label{sect.intro.fa12}

The \href{https://archive.softwareheritage.org/swh:1:cnt:6b3b9423e6d47a813139bc079bc0f53afefa2fd3;origin=https://gitlab.com/nomadic-labs/mi-cho-coq/;visit=swh:1:snp:2e40ba6397bd802a3eaa8ee84ebab385cc070584;anchor=swh:1:rev:f6fbab5abf35945c1ea58a02ffe5c235021b162f;path=/src/contracts_coq/fa12_standard.v}{formal FA1.2 standard} is written in Coq and structured into a small number of modules:
\begin{enumerate*}
\item
\href{https://archive.softwareheritage.org/swh:1:cnt:6b3b9423e6d47a813139bc079bc0f53afefa2fd3;origin=https://gitlab.com/nomadic-labs/mi-cho-coq/;visit=swh:1:snp:2e40ba6397bd802a3eaa8ee84ebab385cc070584;anchor=swh:1:rev:f6fbab5abf35945c1ea58a02ffe5c235021b162f;path=/src/contracts_coq/fa12_standard.v;lines=113}{\lstinline{FA12StorageAccess}}, 
\href{https://archive.softwareheritage.org/swh:1:cnt:6b3b9423e6d47a813139bc079bc0f53afefa2fd3;origin=https://gitlab.com/nomadic-labs/mi-cho-coq/;visit=swh:1:snp:2e40ba6397bd802a3eaa8ee84ebab385cc070584;anchor=swh:1:rev:f6fbab5abf35945c1ea58a02ffe5c235021b162f;path=/src/contracts_coq/fa12_standard.v;lines=141}{\lstinline{FA12StorageDefinitions}}, 
and 
\href{https://archive.softwareheritage.org/swh:1:cnt:6b3b9423e6d47a813139bc079bc0f53afefa2fd3;origin=https://gitlab.com/nomadic-labs/mi-cho-coq/;visit=swh:1:snp:2e40ba6397bd802a3eaa8ee84ebab385cc070584;anchor=swh:1:rev:f6fbab5abf35945c1ea58a02ffe5c235021b162f;path=/src/contracts_coq/fa12_standard.v;lines=185}{\lstinline{FA12StorageAxioms}:}
\quad
These specify internal functions which the smart contract must support (see Figure~\ref{fig.xmpl.1}), along with axioms on their behaviour.
These functions are not entrypoints and cannot be called from outside the smart contract.
They may be explicit in the code of a concrete smart contract implementation, but not necessarily --- e.g. the smart contract might be in a low-level, non-functional language --- so long as they \emph{could} be defined on the contract's underlying data structures.
 
We might call this part of the standard an \deffont{idealised implementation}, where `idealised' is used in the sense of `Platonic ideal', rather than in the sense of `perfect'.
\item
\href{https://archive.softwareheritage.org/swh:1:cnt:6b3b9423e6d47a813139bc079bc0f53afefa2fd3;origin=https://gitlab.com/nomadic-labs/mi-cho-coq/;visit=swh:1:snp:2e40ba6397bd802a3eaa8ee84ebab385cc070584;anchor=swh:1:rev:f6fbab5abf35945c1ea58a02ffe5c235021b162f;path=/src/contracts_coq/fa12_standard.v;lines=233}{\lstinline{FA12Standard}:} 
\quad
This specifies entrypoint behaviour in terms of the functions above
and renders into precise Coq code the English of the FA1.2 standard (whence the module name).
Note however that the formal FA1.2 standard goes beyond the English standard by specifying internal functions rather than just entrypoints as per the previous item, thus it adds some \emph{intensional} content which the English FA1.2 standard lacks.
\item
\href{https://archive.softwareheritage.org/swh:1:cnt:6b3b9423e6d47a813139bc079bc0f53afefa2fd3;origin=https://gitlab.com/nomadic-labs/mi-cho-coq/;visit=swh:1:snp:2e40ba6397bd802a3eaa8ee84ebab385cc070584;anchor=swh:1:rev:f6fbab5abf35945c1ea58a02ffe5c235021b162f;path=/src/contracts_coq/fa12_standard.v;lines=337}{\lstinline{FA12SumOfBalances}:}
\quad
This contains results derived from postulates in the formal FA1.2 standard, so which are guaranteed properties of any FA1.2-compliant smart contract.
\end{enumerate*}

\begin{figure}[th!]
\begin{lstlisting}[basicstyle=\LSTfont]
getAllowance   : data storage_ty -> data address -> data address -> data nat
getBalance     : data storage_ty -> data address -> data nat 
getTotalSupply : data storage_ty -> data nat
setBalance     : data storage_ty -> data address -> data nat     
                                                 -> data storage_ty
setAllowance   : data storage_ty -> data address -> data address ->
                                    data nat     -> data storage_ty
\end{lstlisting}
\caption{Types of key functions from the formal FA1.2 standard}
\label{fig.xmpl.1}
\begin{lstlisting}[basicstyle=\LSTfont]
(** Asking for the balance of an owner we just set yields the new value. *)
Axiom getBalance_setBalance_eq : forall sto owner balance',
  getBalance (setBalance sto owner balance') owner = balance'.

(** Setting a balance leaves everyone else's balances unchanged. *)
Axiom getBalance_setBalance_neq : forall sto owner balance' owner',
  owner <> owner' ->
  getBalance (setBalance sto owner balance') owner' = getBalance sto owner'.
\end{lstlisting}
\caption{Example axioms: \emph{ledger entries are abstract arrays}}
\label{fig.xmpl.2}
\begin{lstlisting}[basicstyle=\LSTfont]
(** Entry point: ep_getBalance *)
Definition ep_getBalance
           (      p : data ep_getBalance_type)
           (    sto : data storage_type      )
           (ret_ops : data (list operation)  )
           (ret_sto : data storage_type      ) :=
  let '(owner, contr) := p in
  let         balance := getBalance sto owner in
  let              op := transfer_tokens nat I balance tokens contr in
    ret_sto = sto /\ ret_ops = [op].
\end{lstlisting}
\caption{Specification for \entrypoint{getBalance} entrypoint}
\label{fig.xmpl.spec}
\begin{lstlisting}[basicstyle=\LSTfont]
(** In the case when the sender is withdrawing from someone else's account,
    they must be authorized to transfer at least the specified amount. *)
(sender <> from -> amount <= getAllowance sto from sender)%N 
\end{lstlisting}
\caption{Translation into Coq of a line from the English standard}
\label{fig.xmpl.authorised}
\begin{lstlisting}[basicstyle=\LSTfont]
Theorem ep_getBalance_sumOfAllBalances
        (    env : @proto_env self_type   )
        (      p : data ep_getBalance_type)
        (    sto : data storage_type      )
        (ret_ops : data (list operation)  )
        (ret_sto : data storage_type      ) :
  ep_getBalance env p sto ret_ops ret_sto ->
  sumOfAllBalances ret_sto = sumOfAllBalances sto.
Proof.
  destruct p as [owner contr]; cbn.
  intros [H_ret_sto _].
  subst; auto.
Qed.
\end{lstlisting}
\caption{Example result: \entrypoint{getBalance} does not affect total supply} 
\label{fig.example.result}

\begin{lstlisting}[basicstyle=\LSTfont]
Definition contract
  := Eval cbv in extract (contract_file_M fa12_camlcase_string.contract 500) I.
\end{lstlisting}
\caption{Parsing a Michelson code string into Mi-Cho-Coq's deep embedding of Michelson}
\label{fig.parsing}
\end{figure}

\begin{example}
Code asserting functions of the idealised implementation is in Figure~\ref{fig.xmpl.1}, and two of its example axioms are in Figure~\ref{fig.xmpl.2} (together, these axioms assert that the ledger is \href{https://en.wikipedia.org/wiki/Array_data_type#Abstract_arrays}{an \emph{abstract array}}).
In these figures, \lstinline{data} is a standard Mi-Cho-Coq~\cite{BernardoCHPT19} wrapper mapping 
\href{https://archive.softwareheritage.org/swh:1:cnt:fa5c55c2ffaf666bd981e45213f32457860cc314;origin=https://gitlab.com/nomadic-labs/mi-cho-coq/;visit=swh:1:snp:2e40ba6397bd802a3eaa8ee84ebab385cc070584;anchor=swh:1:rev:f6fbab5abf35945c1ea58a02ffe5c235021b162f;path=/src/michocoq/semantics.v;lines=313}{(a Coq representation of) Michelson types to Coq types},
and \lstinline{sto} is short for `storage' and represents a state datum (ledger entries, address of admin, total of all balances, and so forth) that is threaded through computations
\end{example}

\begin{remark}
We continue the discussion of the \emph{idealised implementation} above:
The functions in Figure~\ref{fig.xmpl.1} are building blocks with which we can specify the behaviour of the entrypoints listed in Subsection~\ref{subsect.five.entrypoints}.
In this respect, our verification has done something that looks deceptively simple but is not.
By writing these functions down we have refined the English FA1.2 standard --- which speaks only about entrypoints and thus is in some sense purely extensional --- to a specification which is not just more precise (since it is written in Coq); but also adds intensional structure (i.e. not having purely to do with entrypoint behaviour) in that it describes an idealised implementation which a concrete implementation must resemble in a sense made quite formal by the standard itself.
\end{remark}

\begin{example}
\label{example.7}
Code asserting entrypoint behaviour is in \lstinline{FA12Specification}.
For instance:
\begin{enumerate}
\item
Figure~\ref{fig.xmpl.spec} includes code which specifies that a call to the \entrypoint{getBalance} entrypoint should get the balance (this is the \lstinline{balance := getBalance sto owner} part, which is passed to the callback in the operation \lstinline{op}) and any tokens attached to the call just get passed through untouched, as per \href{https://tzip.tezosagora.org/proposal/tzip-5/#getbalance}{a line from the standard} that \emph{``getBalance \dots returns [the] balance of the given address, or zero if no such address is registered.''}.
We spell this out (in small font) in Figure~\ref{fig.closer}: see also \href{https://archive.softwareheritage.org/swh:1:cnt:6b3b9423e6d47a813139bc079bc0f53afefa2fd3;origin=https://gitlab.com/nomadic-labs/mi-cho-coq/;visit=swh:1:snp:2e40ba6397bd802a3eaa8ee84ebab385cc070584;anchor=swh:1:rev:f6fbab5abf35945c1ea58a02ffe5c235021b162f;path=/src/contracts_coq/fa12_standard.v;lines=118-119}{code for \lstinline{getBalance}} and \href{https://archive.softwareheritage.org/swh:1:cnt:6b3b9423e6d47a813139bc079bc0f53afefa2fd3;origin=https://gitlab.com/nomadic-labs/mi-cho-coq/;visit=swh:1:snp:2e40ba6397bd802a3eaa8ee84ebab385cc070584;anchor=swh:1:rev:f6fbab5abf35945c1ea58a02ffe5c235021b162f;path=/src/contracts_coq/fa12_standard.v;lines=147-152}{returning zero if no address is registered}.
\item
Figure~\ref{fig.xmpl.authorised} reflects \href{https://archive.softwareheritage.org/swh:1:cnt:6b3b9423e6d47a813139bc079bc0f53afefa2fd3;origin=https://gitlab.com/nomadic-labs/mi-cho-coq/;visit=swh:1:snp:2e40ba6397bd802a3eaa8ee84ebab385cc070584;anchor=swh:1:rev:f6fbab5abf35945c1ea58a02ffe5c235021b162f;path=/src/contracts_coq/fa12_standard.v;lines=285-287}{in the formal FA1.2 standard}
a condition from the English FA1.2 standard that \emph{``the transaction sender must be previously authorized to transfer at least the requested number of tokens from the \lstinline{"from"} account using the approve entrypoint}.  
\end{enumerate}
\end{example}

\begin{remark}
\label{remark.not.in}
\begin{enumerate}
\item
Continuing Remark~\ref{rmrk.left.out}, we do not assert that an entrypoint call must succeed, even if all of the conditions described in the FA1.2 standard are met, since entrypoints can fail for implementation-specific reasons. 
\item
Furthermore, \lstinline{FA12Specification} is a \emph{ledger standard}, concerned with the conditions for entrypoint calls to succeed, and what happens when they do.
Thus we \emph{do} assert that an entrypoint must fail if conditions for a successful execution are unmet, but we \emph{do not} assert what error it should return and we omit clauses from the English standard of the form \emph{``This entrypoint can fail with the following errors''}. 

This is not because they do not matter (they do, of course) but because from the point of view of the maths of maintaining a ledger, these clauses concern standards for diagnostics and debugging rather than standards for \emph{being a ledger}.
An analogy: it matters if a computation of $100!$ reports \lstinline{NatOverflow} when its 32-bit integers overflow, but not to a mathematical specification of what it is to be \emph{the factorial function}.\footnote{\label{rabbit}%
Error-reporting is also a bit of a rabbit-hole which the English standard skirts but a Coq standard could not.
For example suppose \emph{two} preconditions fail: which associated error should be returned?
The English standard does not say (because it does not care) but from our point of view, disambiguating such issues is a distraction which could also restrict generality, and in ways which would not add value to the standard itself. 
One could even argue that the English FA1.2 standard is in fact two standards intertwined: a ledger standard, and an error-reporting standard, and we have formalised the former.

The Edukera FA1.2 verification \href{https://archive.softwareheritage.org/swh:1:cnt:6a674dd389a0e8c8b42cb8aa51f63cfbb8e6fb96;origin=https://github.com/edukera/archetype-lang;visit=swh:1:snp:1d57f54c27b3b4300155035e6cf732fb5263bf5f;anchor=swh:1:rev:45153bdbc30609d593699ac7854b9ee97a06d042;path=/contracts/fa12.arl;lines=71}{specifies error messages} as per the English standard.
Is this wrong? 
No, just different: their verification follows more in the \emph{verified abstraction of code} style of Remark~\ref{remark.abstraction.of.code} (see also Subsection~\ref{subsect.archetype}) than in the \emph{spec as a logical theory} style of this paper, so for example their specification could just state that the error returned is whatever error \emph{actually is} returned by the implementation they are abstracting.
As always, what we view as a feature depends on what we wish to achieve.}
\end{enumerate}
\end{remark}

\begin{figure}[t]
\begin{lstlisting}[basicstyle=\LSTfonttiny]
let balance := getBalance sto owner in (* `owner` balance retrieved from `sto` and put in `balance` *)
let op := transfer_tokens 
  (* `op` is a transfer_token operation, which will act as a callback to the contract `contr`, sending 
     it the value `balance`. tez transfers and smart contract calls in Tezos are the same thing! *)
  (* Each transfer_token operation has a recipient contract (+ optional entrypoint), an amount of tez, 
     and a parameter.  )
  nat           (* parameter type: each contract (+entrypoint) has a parameter type.
                   In this case, recipient parameter type is `nat`, as it is to receive the `balance`,
                   which is also `nat` *)
  I             (* trivial proof by construction that `nat` is *passable* (technical requirement) *) 
  balance       (* parameter: value sent to `contr`. balance is thus a `nat` *)
  tokens        (* amount of tokens: using the notation `tokens`, we return
                   the number of tez that was sent to this contract and that triggered this execution. 
                   Hence, we just "pass the tokens along". *)
  contr         (* recipient: the contract `contr` will be the receiver of this call. 
                   Note that `contr` comes from the parameter sent to `getBalance`. Thus we have 
                   a "callback" pattern: the value requested is not "returned" to the caller, 
                   instead call back `contr` (which may be the caller but not necessarily) with 
                   the requested value *)
in ret_sto = sto /\ ret_ops = [op]. (* require `op` to be the only returned operation *)
\end{lstlisting}
\caption{Closer look at some code from Figure~\ref{fig.xmpl.spec}}
\label{fig.closer}
\end{figure}

\begin{example}
\label{example.verif}
The module \href{https://archive.softwareheritage.org/swh:1:cnt:6b3b9423e6d47a813139bc079bc0f53afefa2fd3;origin=https://gitlab.com/nomadic-labs/mi-cho-coq/;visit=swh:1:snp:2e40ba6397bd802a3eaa8ee84ebab385cc070584;anchor=swh:1:rev:f6fbab5abf35945c1ea58a02ffe5c235021b162f;path=/src/contracts_coq/fa12_standard.v;lines=337}{\lstinline{FA12SumOfBalances}} contains results valid for any implementation compliant with the formal FA1.2 standard, because they are derived just from postulates of the standard.
Figure~\ref{fig.example.result} illustrates
\href{https://archive.softwareheritage.org/swh:1:cnt:6b3b9423e6d47a813139bc079bc0f53afefa2fd3;origin=https://gitlab.com/nomadic-labs/mi-cho-coq/;visit=swh:1:snp:2e40ba6397bd802a3eaa8ee84ebab385cc070584;anchor=swh:1:rev:f6fbab5abf35945c1ea58a02ffe5c235021b162f;path=/src/contracts_coq/fa12_standard.v;lines=430-442}{one such result}, which 
states that querying a balance does not change the total number of tokens on the ledger, as also returned by \entrypoint{getTotalSupply}. 
This is a relevant result and is also a sanity check on the design, that it postulates enough that this can be proved.
\end{example}

\begin{remark}
We have sketched how the \emph{FA1.2 standard} was refined and formalised into the \emph{formal FA1.2 standard}, which conceptually splits in three parts: 
\begin{enumerate*}
\item
an \emph{idealised implementation}, 
\item
a \emph{behavioural specification} (how external entrypoints are wired to the functions), and 
\item
a small \emph{logical theory} of the idealised implementation and its behaviour, stating in particular that FA1.2-specified entrypoints do not change the total supply of tokens. 
\end{enumerate*}
Next we discuss the workflow of verifying a concrete implementation against the standard. 
\end{remark} 

%%%%%%%%%%%%%%%%%%%
\section{Per-implementation verification}
\label{sect.per-imp}

We have verified three implementations as FA1.2 compliant (see below for what that means): 
\begin{enumerate*}
\item
an \href{https://archive.softwareheritage.org/swh:1:dir:63e41e177cb270b4eaf143cf7253c9b00689c283;origin=https://gitlab.com/camlcase-dev/fa1.2;visit=swh:1:snp:3468f4e76154f9037ee288928c1cccba4e34507e;anchor=swh:1:rev:0cdbf84668b4d5f2a8107aaf78ac8f74d24962d8}{implementation by camlCase} in \href{https://hackage.haskell.org/package/morley}{Morley} (a Haskell eDSL for Michelson contracts),
\item
an \href{https://archive.softwareheritage.org/swh:1:cnt:6a674dd389a0e8c8b42cb8aa51f63cfbb8e6fb96;origin=https://github.com/edukera/archetype-lang;visit=swh:1:snp:1d57f54c27b3b4300155035e6cf732fb5263bf5f;anchor=swh:1:rev:45153bdbc30609d593699ac7854b9ee97a06d042;path=/contracts/fa12.arl}{implementation by Edukera} in \href{https://archetype-lang.org/}{Archetype} (an integrated toolchain for specifying, implementing and verifying Tezos smart contracts), and
\item
a \href{https://archive.softwareheritage.org/swh:1:cnt:7c2fa5012d1e744e91db075ed0c21969a904b7d2;origin=https://gitlab.com/dexter2tz/dexter2tz/;visit=swh:1:snp:74b5f6193e6d831c9ff815c4332e4ea82ce5b44d;anchor=swh:1:rev:1cec9d9333eba756603d6cd90ea9c70d482a5d3d;path=/lqt_fa12.mligo}{liquidity ledger} from the prototype Dexter 2 smart contract by the LIGO lang team, in \href{https://ligolang.org/}{CameLIGO} (a language with ML-like syntax for Tezos smart contracts).
\end{enumerate*}

Concretely, verification proceeds as follows (we consider the camlCase contract):
\begin{enumerate*}
\item
The smart contract is compiled from some high-level smart contract language (Morley, Archetype, CameLIGO\dots), to a Michelson codestring --- Michelson is the low-level stack-based native smart contracts language of the Tezos blockchain --- and stored as a Coq string (see 
\href{https://archive.softwareheritage.org/swh:1:cnt:37763c65f8d814adab96f09532d08d591ce179e4;origin=https://gitlab.com/nomadic-labs/mi-cho-coq/;visit=swh:1:snp:2e40ba6397bd802a3eaa8ee84ebab385cc070584;anchor=swh:1:rev:f6fbab5abf35945c1ea58a02ffe5c235021b162f;path=/src/contracts_coq/fa12_camlcase_string.v}{\lstinline{fa12_camlcase_string.v}}).

Thenceforth we do not work directly with the original source code of the smart contract. 
We may use it for reference, but what gets validated is the Michelson file.\footnote{This is good, in the sense that the Michelson code is what gets executed on-chain.  But note that the Michelson code may be compiler-dependent.  Therefore when we say ``We validated a contract'' this \emph{actually} means ``We validated a Michelson compilation of that contract''.}
\item
This Michelson code string is parsed to a term of Mi-Cho-Coq's deep embedding of Michelson (see \href{https://archive.softwareheritage.org/swh:1:cnt:570ebb8ffbbbd5682451835677350946cc9c2256;origin=https://gitlab.com/nomadic-labs/mi-cho-coq/;visit=swh:1:snp:2e40ba6397bd802a3eaa8ee84ebab385cc070584;anchor=swh:1:rev:f6fbab5abf35945c1ea58a02ffe5c235021b162f;path=/src/contracts_coq/fa12_camlcase.v}{\lstinline{fa12_camlcase.v}}).
This is a \href{https://archive.softwareheritage.org/swh:1:cnt:570ebb8ffbbbd5682451835677350946cc9c2256;origin=https://gitlab.com/nomadic-labs/mi-cho-coq/;visit=swh:1:snp:2e40ba6397bd802a3eaa8ee84ebab385cc070584;anchor=swh:1:rev:f6fbab5abf35945c1ea58a02ffe5c235021b162f;path=/src/contracts_coq/fa12_camlcase.v;lines=51}{one-line operation}; see Figure~\ref{fig.parsing}.

Thus we now have dynamically, in memory a Coq datum \lstinline{contract} representing the Michelson code string read from disk,\footnote{So ``We validated one particular compilation to Michelson of that contract'' acutally means ``We validated a Coq datum representing one particular compilation to Michelson of that contract''.}
of which properties can be asserted and proved.
\item
Details of the concrete implementation --- how data is stored, any additional entrypoints and their behaviour --- are packaged up, abstracted, and proved as high-level descriptions in Coq of behaviour (see \href{https://archive.softwareheritage.org/swh:1:cnt:570ebb8ffbbbd5682451835677350946cc9c2256;origin=https://gitlab.com/nomadic-labs/mi-cho-coq/;visit=swh:1:snp:2e40ba6397bd802a3eaa8ee84ebab385cc070584;anchor=swh:1:rev:f6fbab5abf35945c1ea58a02ffe5c235021b162f;path=/src/contracts_coq/fa12_camlcase.v}{\lstinline{fa12_camlcase.v}}). 
\item
Finally we prove that the high-level description of the implementation satisfies the formal FA1.2 specification (see \href{https://archive.softwareheritage.org/swh:1:cnt:60c6e8bac6e7d3fe21a7d8d94c14a1354b009a45;origin=https://gitlab.com/nomadic-labs/mi-cho-coq/;visit=swh:1:snp:2e40ba6397bd802a3eaa8ee84ebab385cc070584;anchor=swh:1:rev:f6fbab5abf35945c1ea58a02ffe5c235021b162f;path=/src/contracts_coq/fa12_camlcase_verif.v}{\lstinline{fa12_camlcase_verif.v}}).

And thus we conclude that the contract is \emph{FA1.2-compliant}. 
\end{enumerate*}
Let's unpack that.
The sentence \emph{``And thus we conclude that the contract is FA1.2-compliant''} is shorthand for a fuller statement that:
\begin{quote}
\emph{A high-level Coq description of a Mi-Cho-Coq datum representing a Michelson code compilation of the original smart contract, satisfies a Coq formalisation of a refinement of the FA1.2 standard.}
\end{quote}
Let's unpack that further to spell out what parts of this are mathematically assured: 
\begin{enumerate*}
\item
Refining the English FA1.2 standard to the formal FA1.2 standard is not mathematically assured.
This was a creative human step of taking the FA1.2 English description and obtaining from it something formal in Coq that is more intensional, extensive, and precise than the English source, yet which we can still judge to be in some sense faithful to it. 
\item
The compilation of the smart contract from its original source code to Michelson is not assured, unless the compiler is verified in some way --- which currently isn't the case for Morley, Archetype and LIGO.\footnote{Morley is more of a macro language for Michelson, but it includes non-trivial transformation of the source code that are not yet proven to preserve semantics.}
Note also that the Michelson code is what gets executed, which localises any subsequent validation to \emph{that} compilation, and not some other compilation e.g. using a different compiler or a different version of that compiler.
\item
We have to trust correctness of the transformation of the Michelson code string into Mi-Cho-Coq's representation of Michelson code; and that Mi-Cho-Coq itself correctly captures the intended semantics of Michelson.
\item
Everything else is rigorous, provided we trust the Coq kernel.
\end{enumerate*}

\begin{figure}%
\hspace{-1em}\centering
\subfloat[\label{fig.a}\centering Spec as verified abstraction of code]{{
\begin{tikzpicture}[xscale=2,yscale=1, auto,swap]
\foreach \pos/\name/\disp in {
  {(0,0)/1/English Spec},
  {(1,1)/2/Formal Spec},
  {(1,-1)/3/Implementation},
  {(2,0)/4/Verification}}
\node[minimum size=20pt,inner sep=2pt] (\name) at \pos {\disp};
\foreach \source/\dest in {
  1/2,1/3,
  2/4,3/4,3/2}
\path[draw,thick,->] (\source) -- node {} (\dest);
\end{tikzpicture}
}}%
\subfloat[\label{fig.b}\centering Spec as logical theory]{{
\begin{tikzpicture}[xscale=1.6,yscale=.8, auto,swap]
\foreach \pos/\name/\disp in {
  {(0,0)/1/Eng Spec},
  {(1.5,0)/2/Form Spec},
  {(3,0)/3/Theory},
  {(2,1)/i1/Impl 1},
  {(2,-1)/i2/Impl 2},
  {(2,-2)/i3/Impl 3},
  {(4,1)/v1/Verif 1},
  {(4,-1)/v2/Verif 2},
  {(4,-2)/v3/Verif 3}}
\node[minimum size=20pt,inner sep=2pt] (\name) at \pos {\disp};
\foreach \source/\dest in {
  1/2,2/3,
  1/i1,1/i2,1/i3,
  i1/v1,i2/v2,i3/v3,
  3/v1,3/v2,3/v3}
\path[draw,thick,->] (\source) -- node {} (\dest);
\end{tikzpicture}
}}%
\ \\[2ex]
\caption{Two workflow architectures}%
\label{fig.workflow}%
\end{figure}
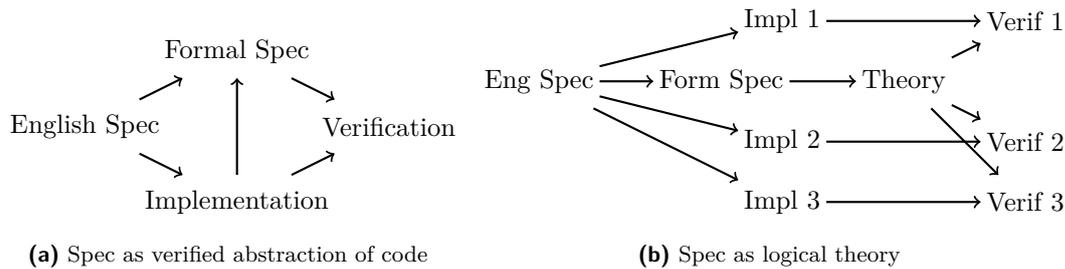

\begin{remark}
\label{remark.abstraction.of.code}
There are (at least) two ways to use logic: to communicate meaning about specific objects (``this chair I'm sitting on, is black''), or to reason about (possibly empty) classes of things (``black chairs''; ``the King of France''). 
Likewise there are two ways to view a formal specification: as a higher-level description of properties of a specific piece of code, or as a specification of properties which pieces of code in general may or may not have. 
This distinction matters for how we write and evaluate the usefulness of our verifications:
\begin{enumerate}
\item
Figure~\ref{fig.a} illustrates one verification workflow.
A programmer reads a specification in English and writes an implementation (bottom left arrow).
Then for additional assurance a formal spec is designed --- in the light of the informal English spec and implementation (top left and middle arrows) --- and the implementation is verified (two right-hand arrows).
Here, \emph{the formal spec is a verified abstraction of the code}.
\item
Figure~\ref{fig.b} illustrates our workflow.
An English specification is refined to some Coq code (the \emph{formal specification}) which entails via its definitions and axioms a collection of properties (\emph{a theory}) against which implementations can be verified.\footnote{This is a kind of dual to \emph{program extraction}, where we start from a specification and extract an executable which then compiles to byte- or machine-code, which (if we trust our compilers) is correct by construction.}
Here implementations are viewed as \emph{models of the spec as a logical theory}. 
If the verifications fail that's an error, and the specification, the implementation, or the theory are modified until they succeed. 
\end{enumerate}
In Figure~\ref{fig.a} we see implementational choices and may even \emph{want} to represent them in the abstract description (a concrete example is in Footnote~\ref{rabbit}, second paragraph).
In Figure~\ref{fig.b} we cannot see implementational choices when we design the abstract description and we do \emph{not want} to.
We will see how this \emph{lack} of access will help us to detect ambiguities in the source English standard in Section~\ref{sect.refining}.
\end{remark}

\begin{remark}
\label{remark.axioms}
We would argue that our workflow in Figure~\ref{fig.b} is a natural way to structure verification against a standard, especially if we plan to verify more than one ledger.
The specification-as-a-theory maximises modularity and reuse, minimises reinventing of the wheel, and accommodates both \emph{a posteriori} and \emph{a priori} reasoning:
\begin{itemize}
\item
\emph{A posteriori.}\ Write your smart contract in whatever language you prefer.
Compile it to Michelson code as you would have to anyway; then (guided by the original source code) \emph{rebuild} a certified correct high-level description of your contract in Coq, 
prove that the certified high-level description satisfies the formal standard, 
and that (the representation in Coq of) the compiled Michelson respects this description.
\item
\emph{A priori.}\ Express a high-level design in Coq (or translate one into Coq). 
Prove it satisfies the formal standard, thus validating your design.
Then implement this design in your language of choice, and 
verify that it respects the high-level description.
\end{itemize}
We would submit to the reader that this is reasonable and that most software development follows some mix of the two patterns above.
\end{remark}

\begin{example}
We continue Example~\ref{example.verif}.
Two typical results in the per-implementation files, which exemplify the kind of results they contain, are that: 
\begin{itemize}
\item
Validity of storage is preserved by all entrypoints.
This is a key sanity property which must include the five entrypoints mentioned in the FA1.2 standard (as listed in Subsection~\ref{subsect.five.entrypoints}) but must also include any other operations offered by the smart contract.
\item
The total supply of tokens is is correctly preserved (or updated, if tokens were minted or burned), and in particular that \entrypoint{getTotalSupply} 
really does return the total supply. 

This is not entirely trivial because, for computational efficiency, most smart contracts track the total number of tokens separately from the tokens themselves.\footnote{An analogy: the Bank of England may keep track of how much cash is in circulation, but it would be computationally prohibitive to actually go out and count all the cash in the country.}
Thus checking that \entrypoint{getTotalSupply}
returns the total supply requires us to write a predicate that computes the total supply, and verify that this `real' total supply is correctly tracked by whatever computationally efficient tally the smart contract is keeping.\footnote{Another analogy: if the reader has ever lost money down the back of a sofa and then struggled (and perhaps failed) to find it again, they may appreciate that making sure that \emph{absolutely no} tokens slip through \emph{any} cracks, may require careful discipline.  Somewhere in the first author's childhood home there may still be a cheque for fifty pounds from his grandfather.}
\end{itemize}
For scale, verification of the first property requires 115 lines of Coq code for the 
%\href{https://gitlab.com/nomadic-labs/mi-cho-coq/-/blob/kristina-fa12-verification-rebase/src/contracts_coq/fa12_camlcase_verification.v#L313}{camlCase contract}, 
\href{https://archive.softwareheritage.org/swh:1:cnt:60c6e8bac6e7d3fe21a7d8d94c14a1354b009a45;origin=https://gitlab.com/nomadic-labs/mi-cho-coq/;visit=swh:1:snp:2e40ba6397bd802a3eaa8ee84ebab385cc070584;anchor=swh:1:rev:f6fbab5abf35945c1ea58a02ffe5c235021b162f;path=/src/contracts_coq/fa12_camlcase_verif.v;lines=488-603}{camlCase contract},
115 lines for the 
%\href{https://gitlab.com/nomadic-labs/mi-cho-coq/-/blob/kristina-fa12-verification-rebase/src/contracts_coq/fa12_edukera_verification.v#L341}{Edukera contract}, 
\href{https://archive.softwareheritage.org/swh:1:cnt:e7ec32030421cdbbeda08766eb72b3d3b86c993d;origin=https://gitlab.com/nomadic-labs/mi-cho-coq/;visit=swh:1:snp:2e40ba6397bd802a3eaa8ee84ebab385cc070584;anchor=swh:1:rev:f6fbab5abf35945c1ea58a02ffe5c235021b162f;path=/src/contracts_coq/fa12_edukera_verif.v;lines=478-593}{Edukera contract},
and 139 for the 
%\href{https://gitlab.com/nomadic-labs/mi-cho-coq/-/blob/kristina-fa12-verification-rebase/src/contracts_coq/fa12_dexter_verification.v#L395}{Dexter 2 contract}.
\href{https://archive.softwareheritage.org/swh:1:cnt:0d81a944bb15a93eb12a6f1c6886cf2983d627d2;origin=https://gitlab.com/nomadic-labs/mi-cho-coq/;visit=swh:1:snp:2e40ba6397bd802a3eaa8ee84ebab385cc070584;anchor=swh:1:rev:f6fbab5abf35945c1ea58a02ffe5c235021b162f;path=/src/contracts_coq/fa12_dexter_verif.v;lines=582-721}{Dexter 2 contract} (roughly half of which is boilerplate code).
\end{example}

\section{Refining the FA1.2 standard}
\label{sect.refining}

FA1.2 is underspecified by design, and often constructively so.
For instance, \href{https://ethtz.io/}{ETHtz}, \href{https://usdtz.com/}{USDtz}, and \href{https://tzbtc.io/}{tzBTC} are all Tezos tokens (wrapping Ether, US Dollars, and Bitcoin respectively), and they may be FA1.2-compliant (or maybe not) --- but clearly they are also different and special.
Being FA1.2-compliant is just a property of a smart contract.
In particular:
\begin{itemize*}
\item 
The standard does not restrict the operations returned by the \entrypoint{transfer} and \entrypoint{approve} entrypoints.
For instance, a contract may call another contract to access its ledger, e.g. if the ledger data is stored remotely.
\item
A contract may have more entrypoints than are mentioned in the standard, e.g. to mint and burn tokens.
\end{itemize*}
However, it is also possible for FA1.2 to be underspecified in undesirable ways, and our verification effort uncovered two such issues, which were updated and corrected:
 
\subsection{Issue 1: Self-transfer}

When the from and to accounts in the \entrypoint{transfer} entrypoint coincide, the operation can be treated either as a NOOP, or as a regular transfer (affecting allowances).
The camlCase implementation originally chose the former; the Edukera and Dexter~2 implementations choose the latter.

It was agreed that this underspecification is undesirable and the FA1.2 standard was \href{https://gitlab.com/tzip/tzip/-/issues/43}{updated to require that this case be treated as a regular transfer}.
The camlCase implementation of the \entrypoint{transfer} entrypoint was \href{https://gitlab.com/camlcase-dev/fa1.2/-/issues/2}{updated accordingly}.

Note how this was noticed because we checked \emph{more than one} ledger implementation against the \emph{same} formal standard (cf. Remark~\ref{remark.abstraction.of.code} and comment~\ref{bloody-minded} of Subsection~\ref{subsect.not.enough}).

\subsection{Issue 2: passing tokens to a view entrypoint}

As noted in Remark~\ref{rmrk.pass.tokens}, when we call an entrypoint in Michelson we must pass it some (possibly zero) number of tez tokens. 
What should an entrypoint do if it gets passed tokens and does not need them?
For instance, the entrypoint could be one of the so-called \deffont{view} entrypoints of FA1.2, \entrypoint{getAllowance}, \entrypoint{getBalance}, and \entrypoint{getTotalSupply}.\footnote{An OO programmer would call the view entrypoints \emph{getters}.}

The camlCase and Edukera implementations opted to keep such tokens.
Thus, if we called camlCase or Edukera implementation of \entrypoint{getBalance} with some tokens, the contract would simply keep the tokens in its balance.
We contacted the creators of the FA1.2 standard and they said this was undesirable: such tokens should be forwarded to the entrypoint's callback --- i.e. a \emph{passthrough}.
The standard was updated to include this condition, and \href{https://gitlab.com/camlcase-dev/fa1.2/-/issues/1}{the implementations updated} accordingly.

\subsection{Summary of refinements}

Thanks to this verification work the FA1.2 standard could be updated to eliminate two missed corner cases.
The implementations were also updated as required.

Notably, the architecture of our verification (as discussed in Section~\ref{sect.per-imp}) had a subtle but powerful effect on the errors that we could detect: 
because of how we factorised our verification files, and because (thanks to this factoring) we could consider \emph{multiple} implementations uniformly against the \emph{same} formal standard, it was easier to see where different implementations had made substantively divergent design decisions and to trace these decisions back to undesirable underspecifications in the core standard.

%%%%%%%%%%%%%%%%%%%%%%%%%%%%%%%%%%%%%%%%%%%
\section{Related and future work}

So far as we know there is nothing else in the literature quite like the FA1.2 formal standard and verifications reported on in this work.
There have however been some other formalisation efforts in this field, notably: the ERC20 standard and its executable semantics in K;
and a formalisation and verification of FA1.2 in Archetype by Edukera.
We discuss each in turn:

\subsection{ERC20-K}

\href{https://eips.ethereum.org/EIPS/eip-20#specification}{ERC20} is to Ethereum as FA1.2 is to Tezos (in fact, ERC20 came first and FA1.2 follows its example).
% https://eips.ethereum.org/EIPS/eip-20#specification
% https://runtimeverification.com/blog/erc20-k-formal-executable-specification-of-erc20/
% https://github.com/runtimeverification/erc20-semantics
% https://gitlab.com/tzip/tzip/-/blob/master/proposals/tzip-7/tzip-7.md
% https://dapphub.github.io/LLsai/token spec in linear logic
% https://github.com/AU-COBRA/ConCert/blob/master/execution/examples/EIP20Token.v 
ERC20 is a quite detailed API specification, but just like the FA1.2 standard, it is written in English, which is neither formal nor executable. 

The \href{http://web.archive.org/web/20210512150957/https://runtimeverification.com/blog/erc20-k-formal-executable-specification-of-erc20/}{ERC20-K semantics} formalises ERC20 in K and annotates it with unit tests, with a particular focus on corner cases.
As per the description:
\begin{quote}
ERC20-K is \dots a formal executable semantics of a refinement of \dots ERC20 [in] the K framework.
ERC20-K clarifies [the precise meaning of] ERC20 functions [and] the corner cases that the ERC20 standard omits \dots such as transfers from yourself to yourself or transfers that result in arithmetic overflows, [and we] manually \dots produced \dots a test-suite [of] tests which we believe cover all the corner cases.
\end{quote}
In other words, ERC20-K turns the English API specification into a executable API specification in K, and provides a detailed test suite of sixty unit tests.

The \href{https://github.com/runtimeverification/erc20-semantics}{ERC20-K homepage} contains references to other work,\footnote{Sadly no published academic work, but see a \href{https://dapphub.github.io/LLsai/token}{linear logic representation by one Rainy McRainface}.} and the broad thrust of its argument is, just like ours, that a standard needs written in a \emph{formal} language.

\subsection{Archetype FA1.2 implementation and verification by Edukera}
\label{subsect.archetype}

The company Edukera has a smart contracts language \emph{Archetype}, in which they wrote a (short and succinct) implementation of an FA1.2-compliant smart contract.
Included with the Archetype source code is \href{https://archive.softwareheritage.org/swh:1:cnt:6a674dd389a0e8c8b42cb8aa51f63cfbb8e6fb96;origin=https://github.com/edukera/archetype-lang;visit=swh:1:snp:1d57f54c27b3b4300155035e6cf732fb5263bf5f;anchor=swh:1:rev:45153bdbc30609d593699ac7854b9ee97a06d042;path=/contracts/fa12.arl;lines=54}{a specification which asserts compliance with the FA1.2 standard}.

In common with our work and with ERC20-K, the development argues for the need for a formal specification against which implementations can be checked.

The verification itself uses a Why3 library for Archetype that implements and specifies Archetype-specific abstractions.
Half of this library is currently verified, which includes the parts that correspond directly to the FA1.2 smart contract, but not all of the libraries on which it depends.\footnote{Details in an \href{http://web.archive.org/web/20210521093913/https://forum.tezosagora.org/t/a-verified-implementation-of-fa1-2/2264}{Agora post}; search for the section on Verification.}
Verification of the rest is a work in progress.

Archetype is an expressive environment in which a user can employ a single set of convenient high-level abstractions to specify and implement a contract, within a uniform and well-automated workflow.\footnote{As per the \href{https://github.com/edukera/archetype-gitbook}{Archetype README}, it provides a \emph{single language to describe [a] business logic \dots from which the different operational versions may be derived.}}
Thus, the Edukera FA1.2 specification is a reflection of the FA1.2 standard into the Archetype toolstack, though as currently written it remains closely-tailored to the sole FA1.2 implementation which it has to talk about, namely the Edukera FA1.2 implementation in Archetype (e.g. if an implementation has additional mint or burn entrypoints, like the Dexter 2 contract, then it will not satisfy the Archetype specification's condition that the total supply is unchanged after each entrypoint).\footnote{%
One could argue that the Archetype FA1.2 specification could be relaxed and the formal FA1.2 standard is also `just' a reflection of the FA1.2 standard into Coq.
This is true but misses the point: it wasn't, because there was not any need, because the camlCase and Dexter 2 contracts do not exist in the Archetype implementation/specification ecosystem, because they are written in Morley and CameLIGO respectively.
The point is not expressivity but scope: Archetype's uniformity and power are available \emph{inside the Archetype toolstack}, whereas you can benefit from the formal FA1.2 standard using any toolstack --- provided you add an entry for a Coq wizard to your budget.  
This is not either/or, but two complementary approaches in a rich design space.} 

By design our work exists at a distance from any specific implementation and indeed from any specific source language, and it can be applied to any contract that can be compiled to Michelson, following a formal standard that does not require the smart contract programmer to buy in to any particular ecosystem except for Tezos itself. 
The correctness guarantee provided by compliance with our formal FA1.2 standard is correspondingly flexible and high-level, and our three verifications (including of the Archetype contract's compilation to Michelson) illustrate how this guarantee can be obtained as part of a practical workflow.

\subsection{Future work}
\label{subsect.future.work}

\paragraph*{Extending to FA2}

The third author is currently extending the development here to the FA2 standard, which is an update and extension of FA1.2 to allow, amongst other things, multiple token types.\footnote{See \href{https://gitlab.com/tzip/tzip/-/blob/master/proposals/tzip-12/tzip-12.md}{Tezos Improvement Proposal 12 (TZIP-12)} and \href{https://web.archive.org/web/20210512095205/https://tezos.b9lab.com/fa2/}{the Tezos Developer Portal FA2 documentation}.}

\paragraph*{Property-based testing}

We argued in Remark~\ref{rmrk.by.construction} above, and in point~\ref{standard.standard} of Subsection~\ref{subsect.not.enough}, that proofs of FA1.2-compliance using our methodology are by construction comparable, because they are all Coq proofs of the same properties --- namely, those stated in the formal FA1.2 standard.

This is true, but not the whole story: what if you have a program and you want to test it?
Here, our development is of little direct help.

Contrast with the Edukera specification and ERC-20K, which come bundled with unit tests which are visibly more portable (we are not aware of the Edukera tests having been made available as a separate portable entity, but the test suite could presumably be ported).

It would be helpful for future work to extend the formal FA1.2 standard to a library of unit tests, or property-based testing properties, against which a prototype smart contract could be plugged, before going to the trouble of running the workflow described in Section~\ref{sect.per-imp}.

\paragraph*{Accessibility}

For sheer accessibility, the work in this paper falls far short of a tool like the \href{https://erc20.fireflyblockchain.com}{ERC20 token verifier}, which will test your bytecode online for compliance with the ERC20 token standard while U wait~\cite{daejun:forvte}, subject to \href{http://web.archive.org/web/20210801144248/https://erc20.fireflyblockchain.com/disclaimer.html}{significant restrictions on the code}.\footnote{For instance: functions not in the ERC20 standard are ignored --- which might sound innocuous but it is not, since without extra functions we might as well use a well-tested smart contract off-the-shelf.  Similarly, the tool does not support external function calls or loops.} 

To the extent that these restrictions map from ERC20 to FA1.2, they do not apply to the work reported in this paper, and we see here the usual trade-off between ease-of-use and power (i.e. between price and performance).
Which we prefer depends on our use case.
 
We could certainly envisage future work in which such a tool is created for FA1.2, based on a test-suite automatically derived from our Coq development.
One could even imagine a tool which inputs a Coq specification (like the formal FA1.2 standard) and outputs an online test-suite, thus combining the rigour of Coq with the accessibility of an online testing suite.
It is early days in this technology and there is much scope for innovation.

%%%%%%%%%%%%%%%%%%%%%%%%%
\section{Conclusion}

Having dependable token ledgers is absolutely necessary for the Tezos blockchain.
Because of the blockchain's modular and updatable architecture, such ledgers are not primitive to the blockchain kernel, and therefore must be coded as smart contracts.\footnote{This is just one small facet of the general fact that innovation in financial technology would benefit from any and all techniques to produce scalable, reliable smart contracts.}

Several ledger implementations exist, both live and deployed (\href{https://ethtz.io/}{ETHtz}, \href{https://usdtz.com/}{USDtz}, and \href{https://tzbtc.io/}{tzBTC}) and prototypical (camlCase, Edukera, and Dexter 2 by the LIGO lang team).

Smart contracts for ledgers are by design intended to handle real value --- and once deployed they may be impossible to change or update. 
Users may lose money if mistakes are made, and also any failures may be perceived as reflecting poorly on the parent Tezos blockchain.\footnote{\dots which may find itself blamed even if the smart contract was created by a third party.}
Therefore, the standards for safety and correctness for this class of program are exceedingly high, not only in the sense that the programs should be right, but also that what `being right' means should be described with clarity and precision.

In particular, it is in the blockchain's best interests that validation of ledger implementations be made as modular as possible, so that proofs and proof-architectures can be reused and presented uniformly and reliably. 

\begin{remark}
\label{rmrk.by.construction}
Before this research, there was an English standard called `the FA1.2 standard', and multiple implementations whose correctness was unknown.
If they were verified (as is the case for the Edukera contract) there was still no way to systematically say what passing that verification meant compared e.g. against another verification by another team working to another interpretation of the English standard.

After this research, we have refined FA1.2 to a precise software artefact in Coq, and verified three implementations against this. 
Thus they are proven \emph{correct}, in \emph{the same way}, with respect to \emph{the same notion of correctness}.

This development is visibly modular and systematic.
Furthermore, the implementations and the standard have both been refined through the detection and elimination of some potentially dangerous corner cases.
We think it can be considered a success. 
\end{remark}

We hope the ideas in this paper may serve as a model for future research and development.

%\bibliography{fmbc-bib}

\end{document}